\documentclass[12pt]{article}
\usepackage[utf8]{inputenc}
\usepackage{hyperref}
\usepackage{amsmath}
\usepackage{graphicx}
\usepackage{enumitem}
\usepackage{enumitem}

\title{\vspace{-2cm}Physical AI Agents: Integrating Cognitive Intelligence with Real-World Action\vspace{1cm}}

\author{
    Fouad Bousetouane\textsuperscript{1,2} \\[1em] 
    \textsuperscript{1}\text{The University of Chicago, USA} \\[0.75em] 
    \textsuperscript{2}\text{2ndsight.ai} \\[1.5em] 
    {\small \href{mailto:bousetouane@uchicago.edu}{\texttt{bousetouane@uchicago.edu}}}
}

\date{}

\begin{document}

\maketitle

\begin{abstract}
Vertical AI Agents are revolutionizing industries by delivering domain specific intelligence and tailored solutions. However, many sectors, such as manufacturing, healthcare, and logistics, demand AI systems capable of extending their intelligence into the physical world, interacting directly with objects, environments, and dynamic conditions. This need has led to the emergence of Physical AI Agents—systems that integrate cognitive reasoning, powered by specialized LLMs, with precise physical actions to perform real-world tasks.

This work introduces Physical AI Agents as an evolution of shared principles with Vertical AI Agents, tailored for physical interaction. We propose a modular architecture with three core blocks—perception, cognition, and actuation—offering a scalable framework for diverse industries. Additionally, we present the Physical Retrieval Augmented Generation \textbf{(Ph-RAG)} design pattern, which connects physical intelligence to industry-specific LLMs for real-time decision-making and reporting informed by physical context.

Through case studies, we demonstrate how Physical AI Agents and the Ph-RAG framework are transforming industries like autonomous vehicles, warehouse robotics, healthcare, and manufacturing, offering businesses a pathway to integrate embodied AI for operational efficiency and innovation.
\end{abstract}

\newpage
\tableofcontents
\newpage

\section{Introduction}

The rapid evolution of Artificial Intelligence (AI) has brought about unprecedented opportunities to address challenges across industries. General-purpose AI systems, while showcasing remarkable capabilities in natural language understanding and task automation, have yet to meet the nuanced demands of highly specialized sectors. The hope for a leap toward Artificial General Intelligence (\textbf{AGI}) remains, but the complexities of precision, adaptability, and dynamic decision-making in real-world environments have shifted the focus toward more targeted solutions.

This shift has given rise to \textbf{Vertical AI Agents}, specialized systems designed to tackle domain-specific challenges with precision and scalability. By leveraging tailored frameworks, these agents are reshaping core software operations and advancing autonomous systems across industries, enabling intelligent decision-making that aligns with unique operational demands.

However, some industries—such as manufacturing, logistics, and healthcare—demand more than cognitive problem-solving. They require AI systems capable of interacting directly with the physical environment, navigating real-world dynamics, and executing precise actions. This necessity has led to the emergence of \textbf{Physical AI Agents}, embodied systems that bridge cognitive reasoning with physical interaction. By seamlessly integrating intelligence into the physical world, Physical AI Agents open new frontiers for AI to address challenges in dynamic and unstructured environments.

\subsection{The Pursuit of AGI: Beyond Generalization}

\textbf{AGI} remains an aspiration \cite{goertzel2007advances},\cite{AGI2}. While the goal is to create AI systems that seamlessly solve problems across industries and domains, no practical solutions currently exist. Research continues to advance, but challenges such as \textbf{data limitations}, \textbf{vertical reasoning gaps}, and \textbf{ethical concerns} have hindered progress. 

General-purpose \textbf{large language models (LLMs)}—such as \textbf{OpenAI GPT-4} \cite{openai2024gpt4technicalreport}, \textbf{Meta LLaMA} \cite{grattafiori2024llama3herdmodels}, and others—represent a significant step in that direction \cite{AGI_GPT}. These models have demonstrated remarkable utility in tasks like natural language understanding, summarization, and even coding assistance, showcasing a level of generalization across multiple domains. However, while LLMs are a notable advancement, they still lack the versatility, adaptability, and contextual awareness required to achieve AGI.

For instance, while general-purpose LLMs have demonstrated significant utility, they often fall short when addressing the nuanced requirements of highly specialized, complex domains. Their inability to provide precise, context-aware solutions becomes evident in scenarios such as:

\begin{itemize}[noitemsep, topsep=0pt] \item In \textbf{supply chain management}, the intricate interplay of logistics, demand forecasting, and global disruptions demands AI systems capable of dynamic decision-making, real-time adaptability, and operational resilience. \item The \textbf{oil and gas} industry presents high-stakes challenges, such as predictive maintenance for drilling equipment and optimizing resource extraction processes, requiring domain-specific insights that go beyond the capabilities of generalist AI. \item In \textbf{healthcare}, tasks like personalized treatment planning, predictive diagnostics, and real-time patient monitoring necessitate systems that are not only sensitive to ethical constraints but also capable of processing highly specialized medical knowledge. \item In \textbf{manufacturing}, dynamic shop floor environments require AI systems capable of handling real-time adaptations, integrating with physical equipment, and ensuring consistent quality control in high-pressure production lines. \end{itemize}

These examples underscore the limitations of general-purpose LLMs in delivering the level of precision, adaptability, and contextual understanding required for these environments. To overcome these barriers, businesses are increasingly turning to \textbf{Vertical AI Agents}—intelligent systems tailored to address domain-specific challenges with unmatched precision and operational relevance.

Nowadays, with the growing need for embodied AI to enhance environmental understanding and enable precise actions and operations, AI is increasingly stepping into the physical realm—AI is getting physical!

The next section explores this critical evolution—how AI is extending into the physical world—and introduces Physical AI Agents as a transformative innovation for industries requiring seamless interaction between cognitive intelligence and real-world adaptability.

\subsection{Bringing AI to the Physical World}

Imagine a world where AI integrates seamlessly into our roads, appliances, vehicles, and workplaces. This transformation is becoming a reality with the emergence of \textbf{Physical AI Agents}. These systems extend AI beyond digital ecosystems, combining cognitive reasoning with sensory perception and precise physical actions to operate dynamically in real-world environments.

Building on the foundation of \textbf{Vertical AI Agents}, which excel in domain-specific decision-making, \textbf{Physical AI Agents} go further by enabling intelligent interaction with and adaptation to physical environments. From navigating complex terrains to performing precise tasks, these agents are reshaping industries where intelligence and physical presence are essential.

The following sections examine the architecture, applications, and transformative potential of \textbf{Physical AI Agents}, highlighting their role in addressing both cognitive and physical challenges with unmatched precision and adaptability.

\subsection{Purpose of This Work}

This work aims to spotlight emerging paradigms in AI that are poised to shape the future of industries and transform daily life. The objectives are:

\begin{enumerate}[noitemsep]
    \item \textbf{Highlight the Need for Vertical Intelligence:}
    \begin{itemize}[noitemsep]
        \item Explore how domain-specific intelligence and agentic systems are revolutionizing industries by enabling precision, adaptability, and operational efficiency.
    \end{itemize}

    \item \textbf{Propose a Standardized Architecture for Physical AI Agents:}
    \begin{itemize}[noitemsep]
        \item Define the core blocks—perception, cognition, and actuation—and establish their foundational overlap with Vertical AI Agents.
        \item Introduce the \textbf{Physical Retrieval-Augmented Generation \\ (Ph-RAG)} design pattern, which connects physical intelligence to industry-specific LLMs for real-time decision-making and context-informed augmentation.
    \end{itemize}
\newpage

    \item \textbf{Showcase Industry Applications:}
    \begin{itemize}[noitemsep]
        \item Provide compelling examples of how Physical AI Agents are being applied in sectors such as manufacturing, healthcare, logistics, and autonomous systems.
    \end{itemize}
\end{enumerate}

By addressing these objectives, this work seeks to bridge the cognitive and physical dimensions of AI, paving the way for transformative innovations across industries.

\section{Vertical AI Agents}

\subsection{Definition}

Vertical AI Agents are specialized intelligent systems designed to solve complex, domain-specific challenges. By leveraging \textbf{domain-specific knowledge} and \textbf{vertical intelligence}, these agents optimize workflows, improve decision-making, and deliver precise, tailored insights. 

At their core, Vertical AI Agents rely on advanced reasoning capabilities powered by a fine-tuned, domain-specialized \textbf{LLM}. Unlike general-purpose AI, these agents are designed to address the nuanced requirements of industries such as \textbf{supply chain management}, \textbf{healthcare}, and \textbf{financial services}.

In our previous work \cite{bousetouane2025}, we extensively explored the paradigm of Vertical AI Agents, providing a comprehensive guide on cross-industry use cases and diverse operational design patterns. For readers interested in a deeper understanding, including tailored frameworks and implementation strategies, we encourage referring to our earlier article.

\subsection{Core Components of Vertical AI Agents}
\begin{figure}[h!]
    \centering
    \includegraphics[width=\textwidth]{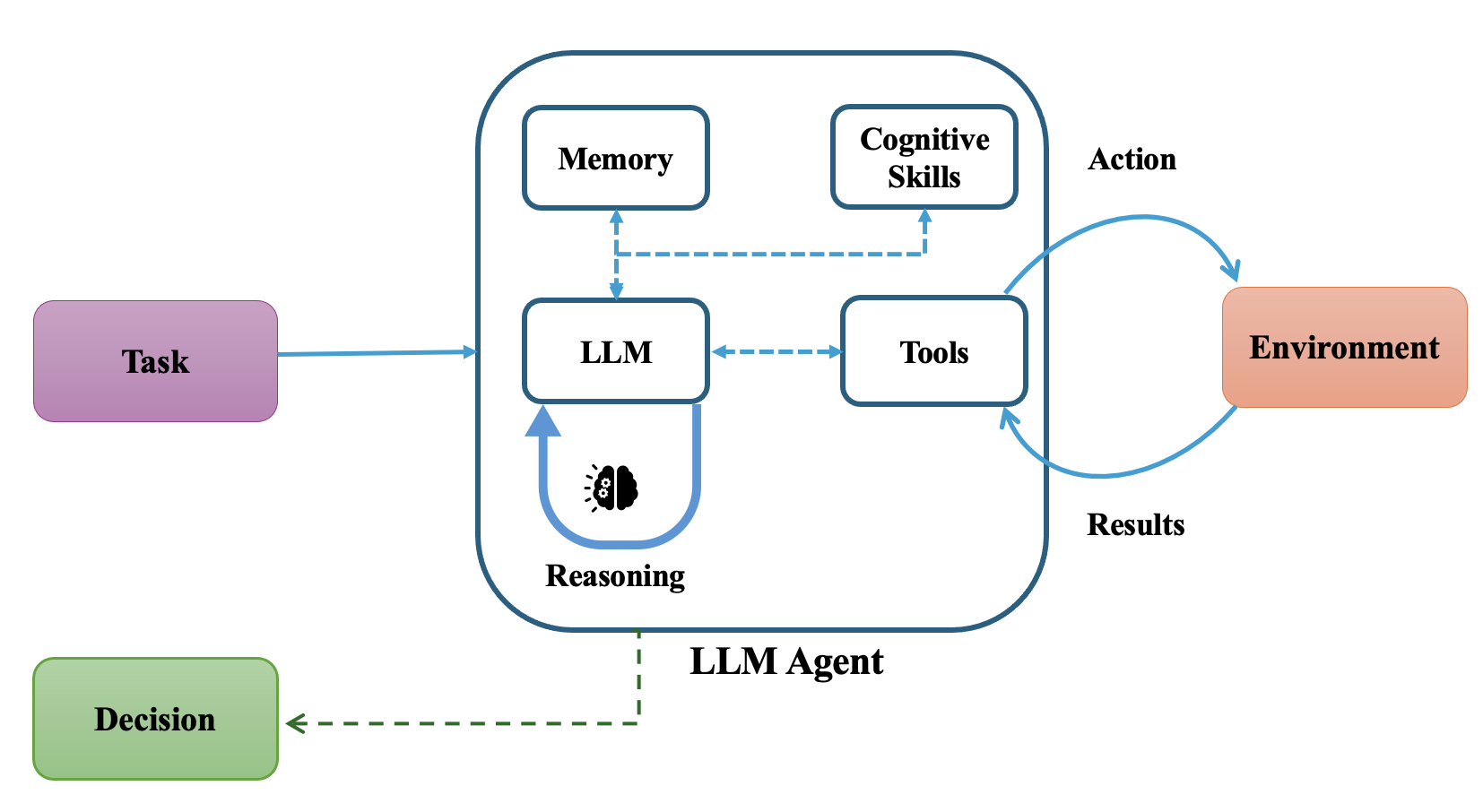}
    \caption{Core Components of a Vertical AI Agent: LLM Backbone, Memory, Cognitive Skills, and Tools \cite{bousetouane2025}.}
    \label{fig:vertical_ai_architecture}
\end{figure}

The architecture of Vertical AI Agents is modular and highly adaptable to domain-specific challenges. The key components, as illustrated in Figure~\ref{fig:vertical_ai_architecture}, include:

\begin{itemize}[noitemsep]
    \item \textbf{Large Language Model (LLM) Backbone:} 
    Provides the foundational reasoning and contextual understanding, fine-tuned to address industry-specific needs.
    \item \textbf{Memory Module:} 
    Retains historical context, past actions, and domain-specific knowledge to enable continuity and adaptive responses.
    \item \textbf{Cognitive Skills Module:} 
    Integrates specialized models for executing precision tasks such as anomaly detection, inventory forecasting, and diagnostics.
    \item \textbf{Tools Module:} 
    Extends the agent’s functionality by connecting it to external systems through vector search, dynamic API integration, and contextual retrieval mechanisms.
\end{itemize}

\subsection{Applications Across Industries}
Below are examples of impactful applications of Vertical AI Agents:
\begin{enumerate}[noitemsep]
    \item \textbf{Supply Chain Management:}
    Vertical AI Agents optimize logistics, streamline inventory management, and predict bottlenecks in real-time. For example, they leverage cognitive skills to forecast demand, recommend replenishment schedules, and dynamically adapt to disruptions in the supply chain.

    \item \textbf{Healthcare:}
    In the healthcare sector, these agents provide predictive diagnostics, enable personalized treatment planning, and improve operational workflows. By integrating medical knowledge with real-time patient data, they support faster, more accurate decision-making and improve patient outcomes.

    \item \textbf{Financial Services:}
    Vertical AI Agents support fraud detection, credit risk assessment, and compliance monitoring. With access to vast financial datasets and domain-specific cognitive skills, they provide insights that enhance risk management and regulatory compliance, ensuring secure and efficient financial operations.
\end{enumerate}

Vertical AI Agents provide the foundation for \textbf{Physical AI Agents}, enabling AI systems to operate in dynamic real-world environments. The \textbf{Cognitive Skills module} bridges this connection by transforming real-world perception data into actionable insights.
The following sections explore the definitions, core components, and real-world applications of \textbf{Physical AI Agents}.

\section{Foundations of Physical AI Agents}
\subsection{Definition}
Physical AI Agents are intelligent, embodied systems designed to interact directly with the physical world. Equipped with advanced sensory capabilities, cognitive intelligence, and precise actuation systems, these agents navigate dynamic environments, manipulate objects, and execute physical tasks with unmatched accuracy and efficiency.

A defining capability of Physical AI Agents is their ability to \textbf{understand and adapt to physical dynamics}, including forces such as \textbf{gravity, friction, and inertia}. This enables them to seamlessly navigate complex environments, handle delicate objects, and perform tasks that require precision and adaptability in real-world conditions.

Beyond understanding physical dynamics, Physical AI Agents are also equipped with \textbf{industry-specific intelligence}, powered by fine-tuned, domain-specific LLMs. These specialized models provide the cognitive backbone for decision-making and contextual understanding, enabling the agent to align its actions with the unique demands of the industry it serves.

By integrating real-time perception, decision-making, and physical execution, Physical AI Agents bridge the gap between digital intelligence and real-world action, offering a transformative solution for tasks that demand physical interaction and domain expertise.

\subsection{Core Components of Physical AI Agents}

Despite their transformative potential, there is no universally accepted architecture for the design and implementation of \textbf{Physical AI Agents}. To address this gap, we present a modular architecture designed to standardize the core components required for these agents across industries and domains.

The proposed architecture is structured around three primary blocks: \textbf{Perception}, \textbf{Cognition}, and \textbf{Actuation}. These blocks are designed to function cohesively, enabling Physical AI Agents to interpret their surroundings, reason and plan based on contextual understanding, and execute precise actions in dynamic environments. By integrating these modular components, the architecture aims to establish a comprehensive framework applicable across various industries and use cases.

\begin{figure}[h!]
    \centering
    \includegraphics[width=\textwidth]{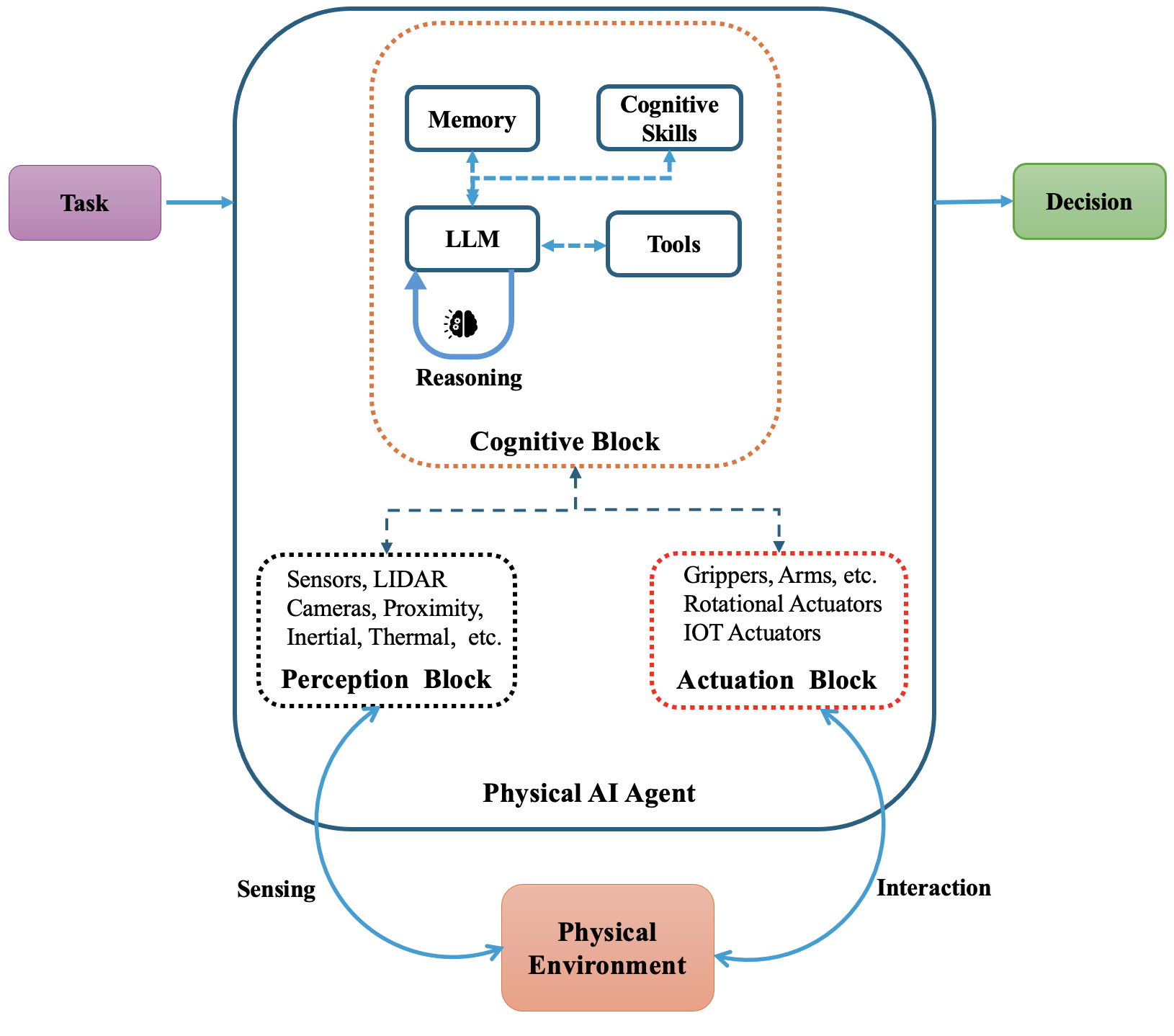}
    \caption{Core Components of a Physical AI Agent: Perception, Cognitive, and Actuation Blocks, with Interaction with the Physical Environment.}
    \label{fig:physical_ai_components}
\end{figure}

Figure~\ref{fig:physical_ai_components} provides an overview of the architecture, highlighting the interactions between the three blocks and the external physical environment.

\begin{enumerate}[noitemsep]
   \item \textbf{Perception Block:}
\begin{itemize}[noitemsep]
    \item The Perception Block is the agent’s sensory interface, enabling it to sense and interpret its surroundings. By capturing and processing real-time environmental data, it provides situational awareness critical for effective decision-making.
    \item \textbf{Sensors:} Includes cameras, LIDAR, proximity sensors, inertial measurement units (IMUs), and IoT-enabled devices.
    \item \textbf{Role:} Captures and processes real-time environmental data, such as spatial layouts, object positions, and motion, providing the agent with situational awareness.
\end{itemize}

\newpage

\item \textbf{Cognitive Block:}
\begin{itemize}[noitemsep]
    \item The Cognitive Block serves as the agent’s reasoning and decision-making hub. It integrates memory, specialized LLM, and cognitive skills to process inputs, plan actions, and adapt to dynamic environments.
    \item \textbf{Role:}
    \begin{itemize}[noitemsep]
        \item Processes sensory data, generates contextual understanding, and translates insights into actionable plans.
        \item Enables real-time decision-making by synthesizing perception data with domain-specific knowledge.
        \item Plans and sequences tasks dynamically based on environmental and operational constraints.
        \item Facilitates adaptability and continuous learning to improve task execution in changing conditions.
    \end{itemize}
    \item \textbf{LLM Reasoning:} A specialized fine-tuned LLM serves as the cognitive backbone, enabling decision-making, contextual understanding, and planning.
    \item \textbf{Memory:} Retains historical context and records past actions, enabling the agent to adapt to changing conditions and maintain continuity.
    \item \textbf{Cognitive Skills:}
    \begin{itemize}[noitemsep]
        \item \textbf{Perception-to-Action Mapping:} Translates sensory inputs into actionable tasks by processing real-time situational data.
        \item \textbf{Understanding Physical Dynamics:} incorporates  models that enable the agent to comprehend physics-related concepts such as gravity, friction, and inertia, allowing for precise navigation, object handling, and real-world interactions.
        \item \textbf{Task-Specific Models:} Executes specialized tasks such as navigation, object recognition, and predictive planning for physical interactions.
    \end{itemize}
    \item \textbf{Tools:} Facilitates integration with external systems such as vector search for domain knowledge and APIs for real-time updates.
\end{itemize}

    \item \textbf{Actuation Block:}
    \begin{itemize}[noitemsep]
        \item The Actuation Block converts cognitive decisions into physical actions. This enables the agent to interact with the environment through precise movements, manipulation, and task execution.
        \item \textbf{Actuators:} Includes robotic arms, grippers, rotational actuators, hydraulic systems, and mobility platforms.
        \item \textbf{Role:} Executes precise physical actions based on decisions from the Cognitive Block, enabling seamless interaction with objects and environments.
    \end{itemize}
\end{enumerate}

\newpage
\subsection{Applications Across Industries}
As Physical AI Agents continue to mature, their versatility and potential to address complex industry challenges become increasingly evident. These systems are reshaping various sectors by seamlessly integrating advanced reasoning, sensory perception, and precise physical actions. Below are real-world examples showcasing their impactful applications:

\begin{enumerate}
    \item \textbf{Autonomous Vehicles}
    \begin{itemize}
        \item \textbf{Scenario:} A self-driving car operates in a busy urban environment, navigating through traffic, pedestrians, and changing road conditions.
        \item \textbf{Solution:} Physical AI Agents leverage an integrated set of components to ensure safe and efficient navigation:
        \begin{itemize}[noitemsep]
            \item \textbf{Cognitive Skills:} Pre-trained models for path planning, collision avoidance, and traffic sign recognition; purpose-built models for motion prediction and real-time decision-making.
            \item \textbf{LLM (Reasoning):} A fine-tuned industry-specific LLM evaluates contextual inputs, generates planning strategies, and ensures coherent and logical decision-making across dynamic traffic scenarios.
            \item \textbf{Tools:} Domain-specific guidance such as dynamic traffic rules, road conditions, and vehicle maintenance policies accessed through vector search or external APIs.
            \item \textbf{Sensors:} LIDAR, cameras, radar, and GPS for real-time environmental perception.
            \item \textbf{Actuators:} Steering, braking, and acceleration systems for precise control and execution.
        \end{itemize}
        \item \textbf{Example Flow:}
        \begin{itemize}[noitemsep]
            \item \textbf{Input:} Sensors detect a pedestrian crossing ahead while the agent receives real-time traffic updates.
            \item \textbf{Perception:} LIDAR and cameras identify the pedestrian’s motion and map the surrounding vehicles.
            \item \textbf{Reasoning:} The LLM evaluates inputs and determines the optimal response using cognitive models for path planning while referencing local traffic rules through Tools.
            \item \textbf{Action:} Actuators apply the brakes to stop safely while maintaining awareness of surrounding traffic.
        \end{itemize}
    \end{itemize}

    \item \textbf{Warehouse Robotics}
    \begin{itemize}
        \item \textbf{Scenario:} A large e-commerce warehouse requires efficient inventory management, including picking, sorting, and replenishing products stored across thousands of bins.
        \item \textbf{Solution:} Warehouse robots combine multiple components to streamline operations:
        \begin{itemize}[noitemsep]
            \item \textbf{Cognitive Skills:} Pre-trained SLAM models for navigation; purpose-built object recognition models for product identification.
            \item \textbf{LLM (Reasoning):} A fine-tuned industry-specific LLM orchestrates task assignment, contextual reasoning, and adaptive planning for efficient workflows.
            \item \textbf{Tools:} Real-time access to inventory systems, order fulfillment policies, and safety protocols through APIs or vector search.
            \item \textbf{Sensors:} Cameras, proximity sensors, and barcode scanners for location tracking and product identification.
            \item \textbf{Actuators:} Robotic arms, grippers, and mobility systems for item retrieval, transport, and replenishment.
        \end{itemize}
        \item \textbf{Example Flow:}
        \begin{itemize}[noitemsep]
            \item \textbf{Input:} The agent receives a request to replenish a specific SKU running low in the system.
            \item \textbf{Perception:} Sensors locate the product and identify its storage bin.
            \item \textbf{Reasoning:} The LLM uses navigation models and inventory policies accessed via Tools to determine the most efficient route and task sequence.
            \item \textbf{Action:} The robot retrieves the product and replenishes the designated shelf.
        \end{itemize}
    \end{itemize}

    \item \textbf{Healthcare Robotics}
    \begin{itemize}
        \item \textbf{Scenario:} A surgical procedure requires a high degree of precision to minimize invasiveness and ensure patient safety.
        \item \textbf{Solution:} Surgical robots integrate advanced components to perform minimally invasive procedures:
        \begin{itemize}[noitemsep]
            \item \textbf{Cognitive Skills:} Pre-trained models for image segmentation and motion stabilization; purpose-built models for tool precision in real-time.
            \item \textbf{LLM (Reasoning):} A fine-tuned medical LLM evaluates surgical plans, patient-specific data, and ongoing sensor inputs to guide robotic actions.
            \item \textbf{Tools:} Access to medical knowledge databases, patient history, and surgical guidelines via APIs or vector search.
            \item \textbf{Sensors:} High-resolution cameras, force sensors, and haptic feedback systems for precise control.
            \item \textbf{Actuators:} Robotic arms with micro-scale precision and surgical tools for executing complex maneuvers.
        \end{itemize}
        \item \textbf{Example Flow:}
        \begin{itemize}[noitemsep]
            \item \textbf{Input:} The agent receives a command to perform a specific incision.
            \item \textbf{Perception:} Sensors provide real-time imaging of the surgical site.
            \item \textbf{Reasoning:} The LLM references surgical guidelines and patient-specific data from Tools to determine the optimal incision path.
            \item \textbf{Action:} The robotic arm performs the incision with precise control and real-time feedback for adjustments.
        \end{itemize}
    \end{itemize}

    \item \textbf{Manufacturing}
    \begin{itemize}
        \item \textbf{Scenario:} An automotive assembly line needs to automate repetitive yet precise tasks such as welding, painting, and quality inspection.
        \item \textbf{Solution:} Smart factory robots integrate advanced components to optimize production:
        \begin{itemize}[noitemsep]
            \item \textbf{Cognitive Skills:} Pre-trained vision models for defect detection; purpose-built models for real-time task sequencing.
            \item \textbf{LLM (Reasoning):} A fine-tuned manufacturing LLM coordinates workflows and optimizes resource allocation in real-time.
            \item \textbf{Tools:} Integration with supply chain data, production guidelines, and maintenance protocols via vector search.
            \item \textbf{Sensors:} Cameras and proximity sensors for defect detection and positioning.
            \item \textbf{Actuators:} Welding arms, painting tools, and conveyor systems for seamless assembly processes.
        \end{itemize}
        \item \textbf{Example Flow:}
        \begin{itemize}[noitemsep]
            \item \textbf{Input:} The agent receives a request to weld specific joints on a car body.
            \item \textbf{Perception:} Cameras and sensors identify the joint's location and assess its condition.
            \item \textbf{Reasoning:} The LLM calculates optimal welding parameters, referencing safety protocols and production standards through Tools.
            \item \textbf{Action:} The robotic arm executes the weld with precision, while quality inspection systems validate the output.
        \end{itemize}
    \end{itemize}

    \item \textbf{Agriculture}
    \begin{itemize}
        \item \textbf{Scenario:} A large-scale farm requires monitoring of crop health, optimization of irrigation, and efficient harvesting.
        \item \textbf{Solution:} Physical AI Agents integrate diverse components to enhance productivity:
        \begin{itemize}[noitemsep]
            \item \textbf{Cognitive Skills:} Pre-trained spectral imaging models for crop health analysis; purpose-built models for irrigation optimization and yield prediction.
            \item \textbf{LLM (Reasoning):} A fine-tuned agricultural LLM analyzes sensor inputs and contextual data to generate actionable insights for farm management.
            \item \textbf{Tools:} Access to historical weather data, soil analysis, and irrigation protocols via APIs or vector search.
            \item \textbf{Sensors:} Drones with multi-spectral cameras, soil moisture sensors, and temperature monitors for environmental monitoring.
            \item \textbf{Actuators:} Autonomous harvesters, irrigation systems, and pesticide sprayers for efficient farm management.
        \end{itemize}

        \newpage
        \item \textbf{Example Flow:}
        \begin{itemize}[noitemsep]
            \item \textbf{Input:} The agent receives a request to optimize irrigation for a specific field.
            \item \textbf{Perception:} Sensors collect data on soil moisture and crop health.
            \item \textbf{Reasoning:} The LLM references irrigation protocols and weather forecasts from Tools to determine the optimal irrigation strategy.
            \item \textbf{Action:} Autonomous irrigation systems adjust water distribution accordingly.
        \end{itemize}
    \end{itemize}
\end{enumerate}

\subsection{Industry Efforts to Build Platforms for Physical AI Agents}

The development of platforms for physical AI agents is in its early stages, with several industry leaders making notable strides. Traditional frameworks like NVIDIA Isaac\cite{nvidia_isaac}, ROS (Robot Operating System)\cite{ros}, AWS RoboMaker\cite{aws_robomaker}, Google Robotics Core\cite{google_robotics}, and Microsoft's robotics solutions\cite{microsoft_robotics} have provided foundational tools for building autonomous agents in conventional ways. These platforms offer essential components for perception, decision-making, and actuation, enabling the creation of autonomous systems.

At CES 2025\cite{ces2025}, NVIDIA unveiled significant advancements aimed at creating truly agentic platforms with generative AI capabilities for reasoning. A highlight is NVIDIA Cosmos\cite{nvidia_cosmos}, a family of foundational AI models designed to train humanoids, industrial robots, and self-driving cars, enhancing their understanding of the physical world through synthetic data generation.

In addition to Cosmos, NVIDIA expanded its Omniverse platform with generative physical AI capabilities, introducing:

\begin{itemize}
    \item \textbf{NVIDIA Edify SimReady:} A generative AI model that can automatically label existing 3D assets with attributes like physics or materials, significantly reducing manual processing time\cite{nvidia_edify}.
    \item \textbf{Omniverse Blueprints:} Designed to accelerate industrial and robotic workflows, these include \textit{Mega} for developing and testing robot fleets at scale within digital twins of industrial environments, and \textit{Autonomous Vehicle (AV) Simulation} for AV developers\cite{omniverse_blueprints}.
\end{itemize}

These developments mark a significant progression toward more advanced, agentic AI systems capable of sophisticated reasoning and interaction within complex physical environments.

\section{Case Study I: Ph-RAG for Oil and Gas Pipeline Integrity Monitoring}

\subsection{Problem Statement}
Monitoring the structural integrity of extensive oil and gas pipelines is a critical task. Failures in pipelines can result in catastrophic environmental damage, operational disruptions, and significant financial losses. Traditional inspection techniques often rely on manual labor and periodic assessments, which are not only resource-intensive but also incapable of providing real-time insights, especially in remote or hazardous locations.

\subsection{Solution Design: \textbf{Ph-RAG} Architecture}
\begin{figure}[h!]
    \centering
    \includegraphics[width=\textwidth]{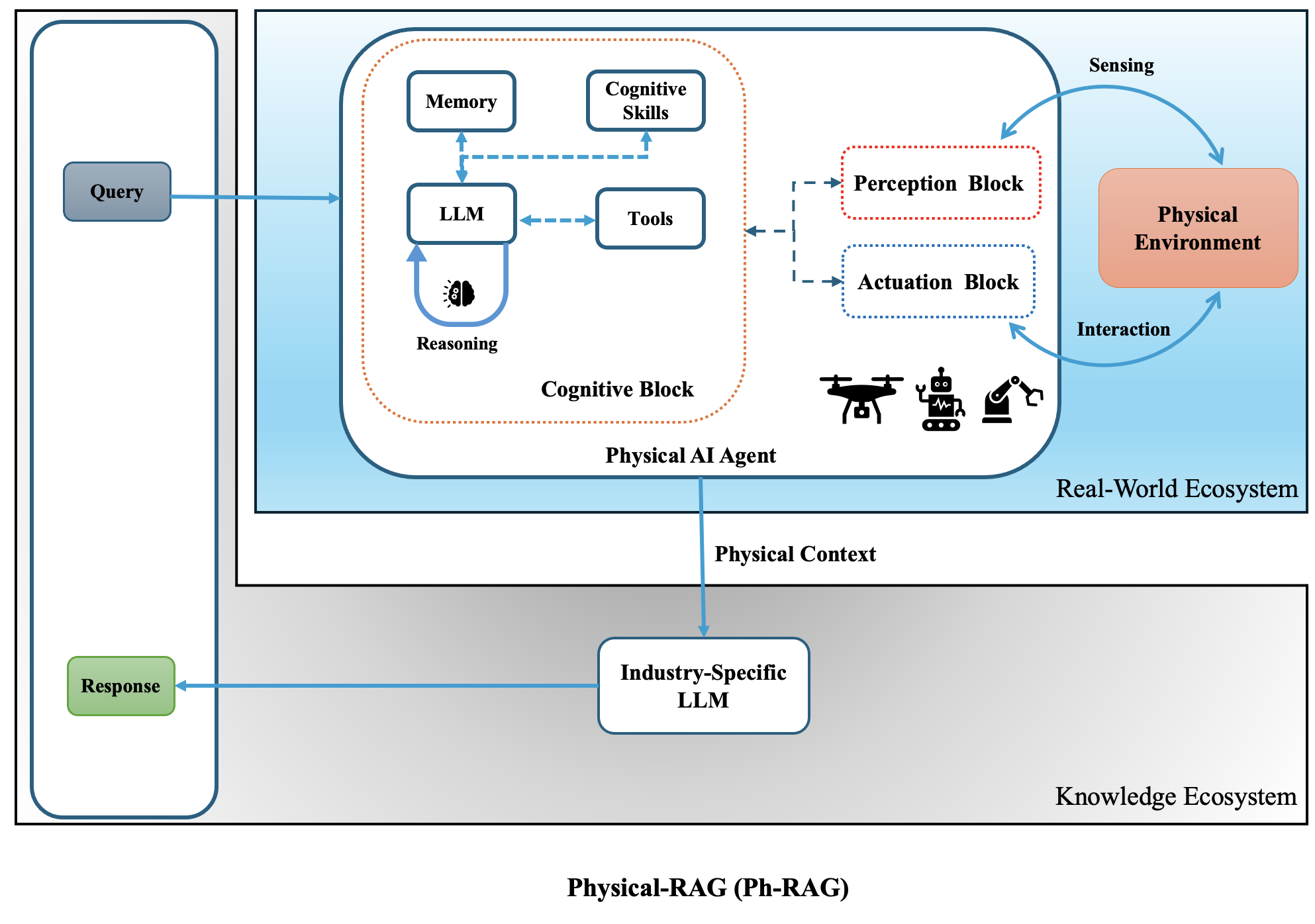}
    \caption{\textbf{Ph-RAG} Architecture - Core Components and Workflow for Pipeline Monitoring.}
    \label{fig:ph_rag_architecture}
\end{figure}

This work introduces the \textbf{Ph-RAG (Physical Retrieval-Augmented Generation)} framework to address these challenges by seamlessly integrating physical intelligence with advanced cognitive reasoning. The architecture, illustrated in Figure~\ref{fig:ph_rag_architecture}, consists of two core components: the \textbf{Physical AI Agent} and an \textbf{industry-specific LLM}.

The \textbf{Physical AI Agent} resides in the \textbf{real-world ecosystem}, performing on-site perception, navigation, and reasoning. It leverages its embedded physical intelligence to adapt to dynamic environments, interact with the physical world, and gather actionable insights. 

The \textbf{industry-specific LLM}, on the other hand, operates within the \textbf{knowledge ecosystem} (digital world), serving as a collaborative partner to the Physical AI Agent. It processes the contextual data provided by the Physical AI Agent to generate comprehensive, domain-specific reports and augment its decision-making capabilities with domain expertise informed by real-world interactions.

\subsection{Components of the Physical AI Agent}
The Physical AI Agent is specifically designed to operate in challenging oil and gas pipeline environments. Its modular architecture enables seamless navigation, data collection, and interaction with the physical world. The key components of the agent include:

\begin{enumerate}[noitemsep]
    \item \textbf{Perception Block:}
    \begin{itemize}[noitemsep]
        \item \textbf{Sensors:} The agent is equipped with high-resolution cameras for visual inspection, LIDAR for structural mapping, acoustic sensors for detecting leaks or vibrations, chemical analyzers for identifying hazardous substances, and thermal imaging units for spotting temperature anomalies.
        \item \textbf{Role:} Processes real-time sensory data to detect corrosion, leaks, structural deformities, and environmental risks. This data forms the foundation for anomaly identification and decision-making.
    \end{itemize}

    \item \textbf{Cognitive Block:}
    \begin{itemize}[noitemsep]
        \item \textbf{Internal Reasoning LLM:} A fine-tuned, task-specific LLM embedded within the agent, responsible for on-site reasoning tasks such as anomaly prioritization, terrain navigation, and pipeline condition assessments.
        \item \textbf{Pre-Trained Cognitive Models:} Specialized models for analyzing pipeline-specific physical dynamics, such as stress distribution, pressure variations, and material degradation.
        \item \textbf{Memory Module:} Stores historical pipeline data, including prior inspections, environmental conditions, and detected anomalies, enabling trend analysis and informed decision-making.
        \item \textbf{Tools Module:} Provides access to operational protocols, environmental regulations, and terrain maps, ensuring the agent adheres to industry standards during inspections.
    \end{itemize}

    \item \textbf{Actuation Block:}
    \begin{itemize}[noitemsep]
        \item \textbf{Actuators:} Includes aerial drones for long-distance and overhead monitoring, ground robots for traversing complex terrains, and modular systems for conducting repairs or deploying advanced sensors.
        \item \textbf{Role:} Enables precise navigation along pipeline routes, execution of inspection tasks, and interaction with physical components to collect detailed data and address critical anomalies.
    \end{itemize}
\end{enumerate}

By combining these components, the Physical AI Agent bridges the gap between real-world environmental understanding, autonomous reasoning, and interaction-driven decision-making.  This architecture, as illustrated in Figure~\ref{fig:ph_rag_architecture}, ensures that actionable insights can be relayed to the industry-specific LLM for generating comprehensive reports, enhancing pipeline integrity monitoring. 

\newpage

\subsection{End-to-End Workflow}
\begin{enumerate}[noitemsep]
    \item \textbf{User Query:} The monitoring team initiates a request for pipeline integrity analysis or sets up alerts for specific anomalies.
    \item \textbf{Physical AI Agent:}
    \begin{itemize}[noitemsep]
        \item Embodied through robotic platforms such as drones or ground-based robots, the Physical AI Agent collects data such as visual, thermal, acoustic, or chemical indicators.
        \item The agent uses its \textbf{internal reasoning LLM} for tasks like navigation, anomaly detection, and planning. This LLM is fine-tuned for reasoning tasks related to real-world conditions, enabling on-the-fly decision-making.
        \item Pre-trained cognitive models process raw sensory data, identify anomalies, and provide structured context for further analysis.
    \end{itemize}
    \item \textbf{Environment Interaction:} The Physical AI Agent adapts to terrain and environmental conditions, ensuring robust performance in challenging settings.
    \item \textbf{Industry-Specific LLM :}
    \begin{itemize}[noitemsep]
        \item The Physical AI Agent sends processed contextual data (e.g., detected anomalies, environmental conditions) to an \textbf{external, industry-specific LLM}, fine-tuned for pipeline monitoring.
        \item The external LLM provides higher-order reasoning, contextualizing the data, identifying trends, and generating actionable insights for the monitoring team.
    \end{itemize}
    \item \textbf{Reporting:} The external LLM delivers detailed reports to the monitoring team, complete with recommendations and identified risks.
\end{enumerate}

This architecture and operational flow are visualized in Figure~\ref{fig:ph_rag_architecture}.

\subsection{Scalability Across Industries}
While this case study focuses on oil and gas pipeline monitoring, the \textbf{Ph-RAG} framework is versatile and scalable. The design pattern can be seamlessly applied to other industries, including infrastructure inspection, environmental monitoring, and precision agriculture, where similar workflows and modular architectures can drive significant operational improvements.

\section{Case Study II: A Hybrid Agentic System for Inventory Management and Product Replenishment}

\subsection{Problem Statement}
Managing inventory and product replenishment in large-scale warehouses is a complex challenge. In a warehouse with over 50,000 products stored in various storage bins and shelves, inefficiencies often arise due to the inability to monitor stock levels in real-time, predict replenishment needs accurately, and execute physical replenishment tasks efficiently. Traditional systems struggle to bridge the gap between virtual inventory management and the physical handling of products, leading to delays, errors, and suboptimal utilization of resources.

\subsection{Solution Design}
A hybrid Agentic AI system integrates the strengths of a \textbf{Vertical AI Agent} and \textbf{Physical AI Agents} to create an intelligent, end-to-end solution. The Vertical AI Agent oversees inventory management, monitors stock levels, and predicts replenishment needs, while Physical AI Agents handle the physical logistics of navigating the warehouse, identifying products, and replenishing shelves. This collaboration minimizes inefficiencies, optimizes workflows, and bridges the digital and physical domains seamlessly.

\subsection{Components of Each AI Agent}

\begin{enumerate}
    \item \textbf{Vertical AI Agent (Inventory Management):}
    \begin{itemize}[noitemsep]
        \item \textbf{Role:} Supervises the inventory system, assigns tasks to Physical AI Agents, and ensures replenishment occurs in real-time.
        \item \textbf{Key Components:}
        \begin{itemize}[noitemsep]
            \item \textbf{Cognitive Skills:} Inventory forecasting model to predict replenishment needs within 30 minutes.
            \item \textbf{Tools:} 
            \begin{itemize}[noitemsep]
                \item Vector search capabilities for domain knowledge retrieval.
                \item Integration with external systems for order processing and fulfillment workflows.
            \end{itemize}
            \item \textbf{Knowledge Domains:}
            \begin{itemize}[noitemsep]
                \item Product-level inventory knowledge.
                \item Historical sales and demand patterns.
                \item Supply chain and fulfillment workflows.
            \end{itemize}
            \item \textbf{Reasoning Module:}
            \begin{itemize}[noitemsep]
                \item Specialized LLM trained to leverage domain knowledge for decision-making and task prioritization.
            \end{itemize}
            \item \textbf{Memory Module:}
            \begin{itemize}[noitemsep]
                \item Stores the agent's past actions, including task assignments and replenishment decisions.
                \item Tracks the history of product movements and operational workflows for continuity.
            \end{itemize}
        \end{itemize}
    \end{itemize}
    
    \item \textbf{Physical AI Agents (Robots):}
    \begin{itemize}[noitemsep]
        \item \textbf{Role:} Execute physical replenishment tasks in the warehouse.
        \item \textbf{Key Components:}
        \begin{itemize}[noitemsep]
            \item \textbf{Sensors:} Cameras and LIDAR for perception and navigation.
            \item \textbf{Cognitive Skills:}
            \begin{itemize}[noitemsep]
                \item SLAM (Simultaneous Localization and Mapping) for warehouse navigation.
                \item Object recognition for identifying products, shelves, and employees.
            \end{itemize}
            \item \textbf{Tools:}
            \begin{itemize}[noitemsep]
                \item Navigation knowledge through vector search for spatial information.
                \item Integration with external systems to ensure efficient item handling and safety compliance.
            \end{itemize}
            \item \textbf{Actuators:} Grippers for picking and placing items.
            \item \textbf{Reasoning Module:}
            \begin{itemize}[noitemsep]
                \item Specialized LLM trained to combine navigation knowledge, safety protocols, and product information for real-time decision-making.
            \end{itemize}

            \newpage
            \item \textbf{Memory Module:}
            \begin{itemize}[noitemsep]
                \item Stores past navigation routes, completed tasks, and replenishment actions.
                \item Maintains a history of operational adjustments and interactions with the environment.
            \end{itemize}
        \end{itemize}
    \end{itemize}
\end{enumerate}

\subsection{End-to-End Flow in Action} 
\begin{enumerate}[noitemsep]
    \item The \textbf{Vertical AI Agent} continuously monitors product stock levels and detects that certain SKUs are approaching low thresholds. Using its inventory forecasting model, it predicts which products will need replenishment in the next 30 minutes.
    \item Based on the replenishment requirements, the Vertical AI Agent assigns tasks to \textbf{Physical AI Agents}, including specific SKUs to retrieve and their corresponding shelf locations.
    \item The \textbf{Physical AI Agents} navigate the warehouse using SLAM, perceiving their environment with cameras and LIDAR while avoiding obstacles and ensuring safety.
    \item Upon reaching the designated storage location, the robots use object recognition to identify the correct products and shelves, then pick up the required items using their grippers.
    \item The robots transport the products to the replenishment location and accurately place them on the shelves.
    \item The Vertical AI Agent updates inventory levels in real-time, ensuring alignment between the digital and physical systems and preparing for future replenishment tasks.
\end{enumerate}

This hybrid system exemplifies the power of combining Vertical and Physical AI Agents to bridge the digital and physical domains. By integrating cognitive intelligence for inventory management with embodied intelligence for physical execution, the system optimizes workflows, minimizes inefficiencies, and transforms traditional warehouse operations into a scalable, adaptive, and highly efficient process.

\section{Conclusion and Future Directions}

\subsection{Conclusion}
This work introduced Physical AI Agents as a transformative paradigm, extending the capabilities of Vertical AI Agents into the physical world. By embedding advanced cognitive reasoning, environmental understanding, and autonomous decision-making into physical platforms, these agents represent a pivotal step toward bridging the gap between digital intelligence and real-world interaction. 

We explored the architecture of Physical AI Agents, defining their core blocks—Perception, Cognition, and Actuation—and demonstrated their adaptability across various industries through real-world use cases. The proposed Physical-RAG (Ph-RAG) design pattern exemplifies how physical intelligence can seamlessly collaborate with industry-specific LLMs to provide real-time, context-informed insights, enhancing operational efficiency and decision-making. 

By integrating modular design principles with domain-specific intelligence, Physical AI Agents redefine how businesses approach complex challenges in dynamic environments. Their ability to process, reason, and act within physical contexts positions them as foundational tools for industries aiming to achieve greater efficiency, scalability, and precision in their operations.

The transformative potential of Physical AI Agents lies not only in their technical capabilities but also in their ability to adapt to diverse environments and integrate seamlessly into existing workflows. As industries continue to evolve, these agents will play a critical role in driving innovation, enabling businesses to meet the demands of an increasingly complex and interconnected world.

\subsection{Future Directions}

To unlock the transformative potential of Physical AI Agents and catalyze innovation at the intersection of research and industry, several strategic priorities must be addressed:

\begin{itemize} \item \textbf{Advancing Physical Intelligence:} Focus on designing advanced cognitive frameworks that integrate real-time perception, spatial reasoning, and adaptive decision-making, enabling agents to perform complex tasks in dynamic environments. \item \textbf{Developing Standardized Architectures:} Establish interoperable and scalable design standards that facilitate seamless integration across diverse industrial applications, fostering ecosystem-wide compatibility. 

\item \textbf{Enhancing Large Language Model (LLM) Synergy:} Augment industry-specific LLMs to process multimodal inputs from physical contexts, enabling precise synthesis and actionable insights tailored to specialized use cases. 

\item \textbf{Optimizing LLMs for Robotic Platforms:} Focus on LLM optimization and quantization techniques to ensure efficient deployment on resource-constrained robotic platforms. This includes reducing computational overhead, minimizing latency, and maintaining accuracy for real-time decision-making. 

\item \textbf{Prioritizing Sustainability and Efficiency:} Pursue research and development in energy-efficient hardware and algorithms to align agent operations with global sustainability initiatives while reducing operational costs. 

\item \textbf{Enabling Multi-Agent Ecosystems:} Investigate cooperative multi-agent systems that enable Physical AI Agents to collaborate in performing large-scale, interdependent tasks across industries such as logistics, healthcare, and manufacturing. \end{itemize}

Addressing these key areas will propel the evolution of Physical AI Agents, bridging the gap between digital intelligence and real-world functionality. These advancements promise to redefine industry standards, drive technological progress, and create unprecedented opportunities for embodied intelligence across global markets.

\newpage
\section{Disclaimer} The architectures, diagrams, and concepts presented in this work are the intellectual property of the author. Proper attribution to the author and this publication is expected when referencing or sharing these materials.

\bibliographystyle{plain} 
\bibliography{science_template} 

\end{document}